\documentclass[runningheads]{svmult}
\usepackage{makeidx}   % allows index generation
\usepackage{subeqnar}  % subnumbers individual equations
\usepackage{physprbb}  % centered layout of diverse elements, etc.
\makeindex             % used for the subject index
\newcommand{\ga}{>}
\newcommand{\la}{<} 
\newcommand{\msun}{M_\odot}
\newcommand{\lya}{Ly$\alpha$~} 
\newcommand{\kel}{~{\rm K}}
\newcommand{\arcdeg}{^\circ}

\begin{document}

\title*{Novel Ways to Probe the Universe with Gamma-Ray Bursts and Quasars}

\toctitle{Novel Ways to Probe the Universe with Gamma-Ray Bursts and
Quasars} 
\titlerunning{Novel Ways to Probe the Universe with Gamma-Ray
Bursts and Quasars}

\author{Abraham Loeb}
\authorrunning{Avi Loeb}
\institute{Astronomy Department, Harvard University\\
Cambridge, MA 02138, USA}

\maketitle              % typesets the title of the contribution

\begin{abstract}

I consider novel ways by which Gamma-Ray Bursts (GRBs) and quasars can be
used to probe the universe. Clues about how and when was the intergalactic
medium ionized can be read off the UV emission spectrum of GRB explosions
from the first generation of stars. The existence of intergalactic and
galactic stars can be inferred from their gravitational microlensing effect
on GRB afterglows.  Prior to reionization, quasars should be surrounded by
a halo of scattered Ly$\alpha$ radiation which probes the neutral
intergalactic medium (IGM) around them. The situation is analogous to the
appearance of a halo of scattered light around a street lamp which is
embedded in a dense fog. Outflows from quasars magnetize the IGM at all
redshifts. As a result, the shocks produced by converging flows during the
formation of large scale structure in the IGM, accelerate electrons to
relativistic energies and become visible in the radio regime through their
synchrotron emission and in the $\gamma$--ray regime through their
inverse--Compton scattering of the microwave background photons. During
transient episodes of strong mergers, X-ray clusters should therefore
appear as extended radio or $\gamma$--ray sources on the sky.

\end{abstract}

\section{Novel Cosmological Studies with Gamma-Ray Bursts}

\subsection{Preface}

Since their discovery four decades ago, quasars have been used as powerful
lighthouses which probe the intervening universe out to high redshifts,
$z\sim 6$ \cite{Hartwick,Fan}.  The spectra of almost all quasars show
strong emission lines of metals, indicating super-solar enrichment of the
emitting gas \cite{Hamann}. This implies that at least in the cores of
galaxies, formation of massive stars and their evolution to supernovae
preceded the observed quasar activity.  If Gamma-Ray Bursts (GRBs)
originate from the remnants of massive stars (such as neutron stars or
black holes), as seems likely based on recent estimates of their energy
output \cite{WKF,Freedman,Frail}, then they should exist at least out to
the same redshift as quasars.  Although GRBs are transient events, their
peak optical-UV flux can be as bright as that of quasars. Hence, GRBs
promise to be as useful as quasars in probing the high--redshift universe.

\begin{figure}[htbp]
\includegraphics{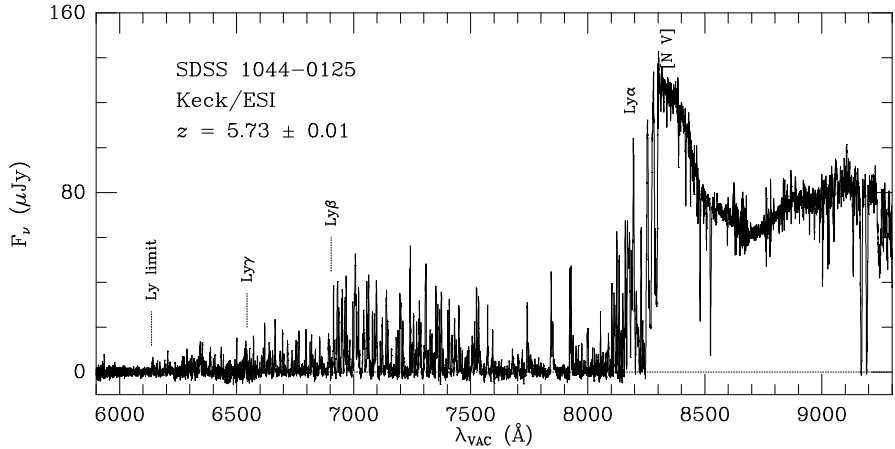}
\vspace{2.8in}
\caption{Spectrum of the quasar SDSS1044-0125 at $z=5.73$ (Djorgovski et
al. 2001), originally discovered by the Sloan Digital Sky Survey (Fan et
al. 2000). }
\label{fig1}
\end{figure}

Not much is known observationally about the universe in the redshift
interval $z=6$--$30$, when the first generation of galaxies condensed out
of the primordial gas left over from the Big Bang (see reviews
\cite{Barkana} and \cite{Loeb_Barkana}). Observations of the microwave
background anisotropies indicate that the cosmic gas became neutral at
$z\sim 1000$ and remained so at least down to $z\sim 30$ (see,
e.g. \cite{Wang}).  On the other hand, the existence of transmitted flux
shortward of the Ly$\alpha$ resonance in the spectrum of the
highest-redshift quasars and galaxies (see, for example,
Figure~\ref{fig1}), indicates that the intergalactic medium was reionized
to a level better than 99.9999\% by a redshift $z\sim 6$.  This follows
from the fact that the Ly$\alpha$ optical depth of the intergalactic medium
at high-redshifts ($z\gg1$) is \cite{Gunn},
\begin{equation}
\tau_{\alpha}={\pi e^2 f_\alpha \lambda_\alpha n_{HI}(z) \over m_e
cH(z)} \approx 6.15\times 10^5 x_{HI} \left({\Omega_bh\over
0.03}\right)\left({\Omega_m\over 0.3}\right)^{-1/2} \left({1+z\over
10}\right)^{3/2}, 
\label{eq:G-P}
\end{equation}
where $H\approx 100h~{\rm km~s^{-1}~Mpc^{-1}}\Omega_m^{1/2}(1+z)^{3/2}$ is
the Hubble parameter at the source redshift $z$, $f_\alpha=0.4162$ and
$\lambda_\alpha=1216$\AA\, are the oscillator strength and the wavelength
of the Ly$\alpha$ transition; $n_{HI}(z)$ is the average intergalactic
density of neutral hydrogen at the source redshift (assuming primordial
abundances); $\Omega_m$ and $\Omega_b$ are the present-day density
parameters of all matter and of baryons, respectively; and $x_{HI}$ is the
average fraction of neutral hydrogen. Modeling \cite{Fan} of the {\it
transmitted flux} in Figure~\ref{fig1} implies $\tau_{\alpha}<0.5$ or
$x_{HI}< 10^{-6}$, i.e., most of the low-density gas throughout the
universe is ionized to a level of 99.999\% at $z\la 6$. However, there are
some dark intervals in the spectrum which could be indicative of regions
with a higher neutral fraction \cite{Djorgovski}. In fact, Becker et
al. \cite{Becker} reported the detection of an extended ($\ga 300$\AA~
long) dark interval just shortward of the Ly$\alpha$ emission line in the
spectrum of the newly discovered quasar SDSS 1030+0524 at $z=6.28$.  The
suppression of the flux by a factor $\ga 150$ may indicate the first
detection of the Gunn-Peterson trough, although caution is warranted since
the inferred optical depth $\tau_\alpha\ga 5$ can be produced by a neutral
fraction as small as $X_{HI}\sim 2 \times 10^{-4}$ according to
equation~(\ref{eq:G-P}).

The question: {\it how and when was the universe reionized?}  defines a new
frontier in observational cosmology \cite{Loeb_Barkana}. The UV spectrum of
GRB afterglows can be used to probe the ionization and thermal state of the
intergalactic gas during the epoch of reionization, at redshifts $z\sim
7$--10 \cite{Miralda_Escude}.  The stretching of the temporal evolution of
GRB lightcurves by the cosmological redshift factor $(1+z)$, makes it
easier for an observer to react in time and measure a spectrum of their
optical-UV emission at its peak.

Energy arguments suggest that reionization resulted from photoionization
and not from collisional ionization \cite{Tegmark,Furlanetto}.  The
corresponding sources of UV photons were either stars or quasars.  Recent
simulations of the first generation of stars that formed out of the
primordial metal--free gas indicate that these stars were likely massive
\cite{Bromm,Abel}. If GRBs result from compact stellar remnants, such as
black holes or neutron stars, then the fraction of all stars that lead to
GRBs may have been higher at early cosmic times. This, however, is true
only if the GRB phenomena is triggered on a time scale much shorter than
the age of the universe at the corresponding redshift, which for $z\gg 1$
is $\sim 5.4\times 10^8~{\rm yr}~(h/0.7)^{-1}(\Omega_m/0.3)^{-1/2}
[(1+z)/10]^{-3/2}$. 
%This condition may not hold, for example, for neutron
%star binaries with an excessively long coalescence time.

\subsection{Properties of High-Redshift GRB Afterglows}

\begin{figure}[htbp]
\includegraphics{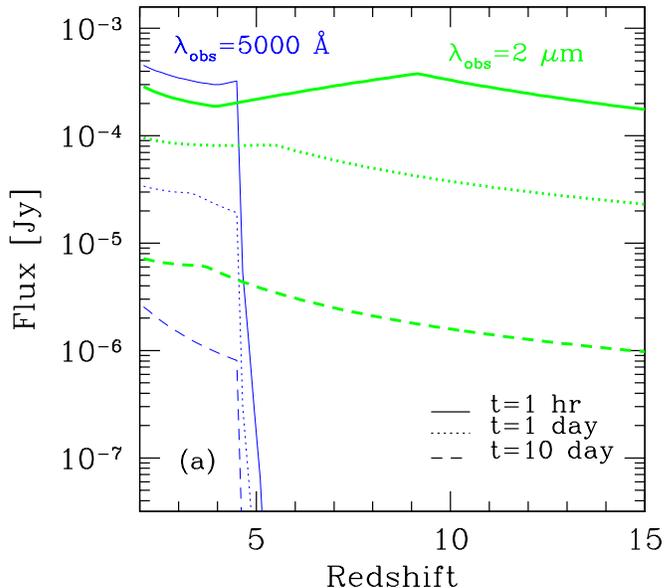}
\vspace{3.25in}
\caption{Theoretical expectation for the observed afterglow flux of a GRB
as a function of its redshift (from Ciardi \& Loeb 2000).  The curves refer
to an observed wavelength of $5000$ \AA (thin lines) and $2 \mu$m (thick
lines). Different line types refer to different observed times after the
GRB trigger, namely 1 hour (solid line), 1 day (dotted) and 10 days
(dashed). The $5000$\AA~ flux is strongly absorbed at $z>4.5$ by
intergalactic hydrogen. However, at infrared and radio wavelengths the
observed afterglow flux shows only a mild dependence on the source
redshift.}
\label{fig2}
\end{figure}

Young (days to weeks old) GRBs outshine their host galaxies in the optical
regime. In the standard hierarchical model of galaxy formation, the
characteristic mass and hence optical luminosity of galaxies and quasars
declines with increasing redshift \cite{Haiman,Stern,Barkana}. Hence, GRBs
should become easier to observe than galaxies or quasars at increasing
redshift. Similarly to quasars, GRB afterglows possess broad-band spectra
which extend into the rest-frame UV and can probe the ionization state and
metallicity of the IGM out to the epoch when it was reionized at redshifts
$z\sim 7$--$10$ \cite{Loeb_Barkana}.  Lamb \& Reichart \cite{Lamb} have
extrapolated the observed $\gamma$-ray and afterglow spectra of known GRBs
to high redshifts and emphasized the important role that their detection
might play in probing the IGM. Simple scaling of the long-wavelength
spectra and temporal evolution of afterglows with redshift implies that at
a fixed time lag after the GRB trigger in the observer's frame, there is
only a mild change in the {\it observed} flux at infrared or radio
wavelengths as the GRB redshift increases. Ciardi \& Loeb \cite{Ciardi}
demonstrated this behavior using a detailed extrapolation of the GRB
fireball solution into the non-relativistic regime (see the 2$\mu$m curves
in Figure~\ref{fig2}). Despite the strong increase of the luminosity
distance with redshift, the observed flux for a given observed age is
almost independent of redshift in part because of the special spectrum of
GRB afterglows (see Figure~\ref{fig4}), but mainly because afterglows are
brighter at earlier times and a given observed time refers to an earlier
intrinsic time in the source frame as the source redshift increases. The
mild dependence of the long-wavelength ($\lambda_{\rm obs}>1\mu$m) flux on
redshift stands in contrast to other high-redshift sources such as galaxies
or quasars, which fade rapidly with increasing redshift
\cite{Haiman,Stern,Barkana}. Hence, GRBs provide exceptional lighthouses
for probing the universe at $z=6$--30, at the epoch when the first stars
had formed.

Assuming that the GRB rate is proportional to the star formation rate and
that the characteristic energy output of GRBs is $\sim 10^{52}~{\rm ergs}$,
Ciardi \& Loeb \cite{Ciardi} predicted that there are always $\sim 15$ GRBs
from redshifts $z> 5$ across the sky which are brighter than $\sim 100$ nJy
at an observed wavelength of $\sim 2\mu$m.  The infrared spectrum of these
sources could be taken with future telescopes such as the {\it Next
Generation Space Telescope} (planned for launch in 2009; see
http://ngst.gsfc.nasa.gov/), as a follow-up on their early X-ray
localization with the {\it Swift} satellite (planned for launch in 2003;
see http://swift.sonoma.edu/).

The redshifts of GRB afterglows can be estimated photometrically from
either the Lyman limit or Ly$\alpha$ troughs in their spectra. At low
redshifts, the question of whether the Lyman limit or Ly$\alpha$ trough
interpretation applies depends on the absorption properties of the host
galaxy. If the GRB originates from within the disk of a star--forming
galaxy, then the afterglow spectrum will likely show a damped Ly$\alpha$
trough. At $z> 6$ the Ly$\alpha$ trough would inevitably exist since the
intergalactic Ly$\alpha$ opacity is $> 90\%$ (see Figure 13 in
\cite{Stern}). Interestingly, an absorption feature in the afterglow
spectrum which is due to the neutral hydrogen within a molecular cloud or
the disk of the host galaxy, is likely to be time-dependent due to the
ionization caused by the UV illumination of the afterglow itself along the
line-of-sight \cite{Perna}.

So far, there have been two claims for high-redshift GRBs. Fruchter
\cite{Fruchter} argued that the photometry of GRB 980329 is consistent with
a Ly$\alpha$ trough due to IGM absorption at $z\sim 5$.  Anderson et
al. \cite{Anderson} inferred a redshift of $z=4.5$ for GRB 000131 based on
a crude optical spectrum that was taken by the VLT a few days after the GRB
trigger. {\it These cases emphasize the need for a coordinated observing
program that will alert 10-meter class telescopes to take a spectrum of an
afterglow about a day after the GRB trigger, based on a photometric
assessment (obtained with a smaller telecope using the Lyman limit or
Ly$\alpha$ troughs) that the GRB may have originated at a high redshift.}

In the following two subsections, I illustrate the usefulness of GRB
afterglows for cosmological studies through two examples.

\begin{figure}[htbp]
\includegraphics{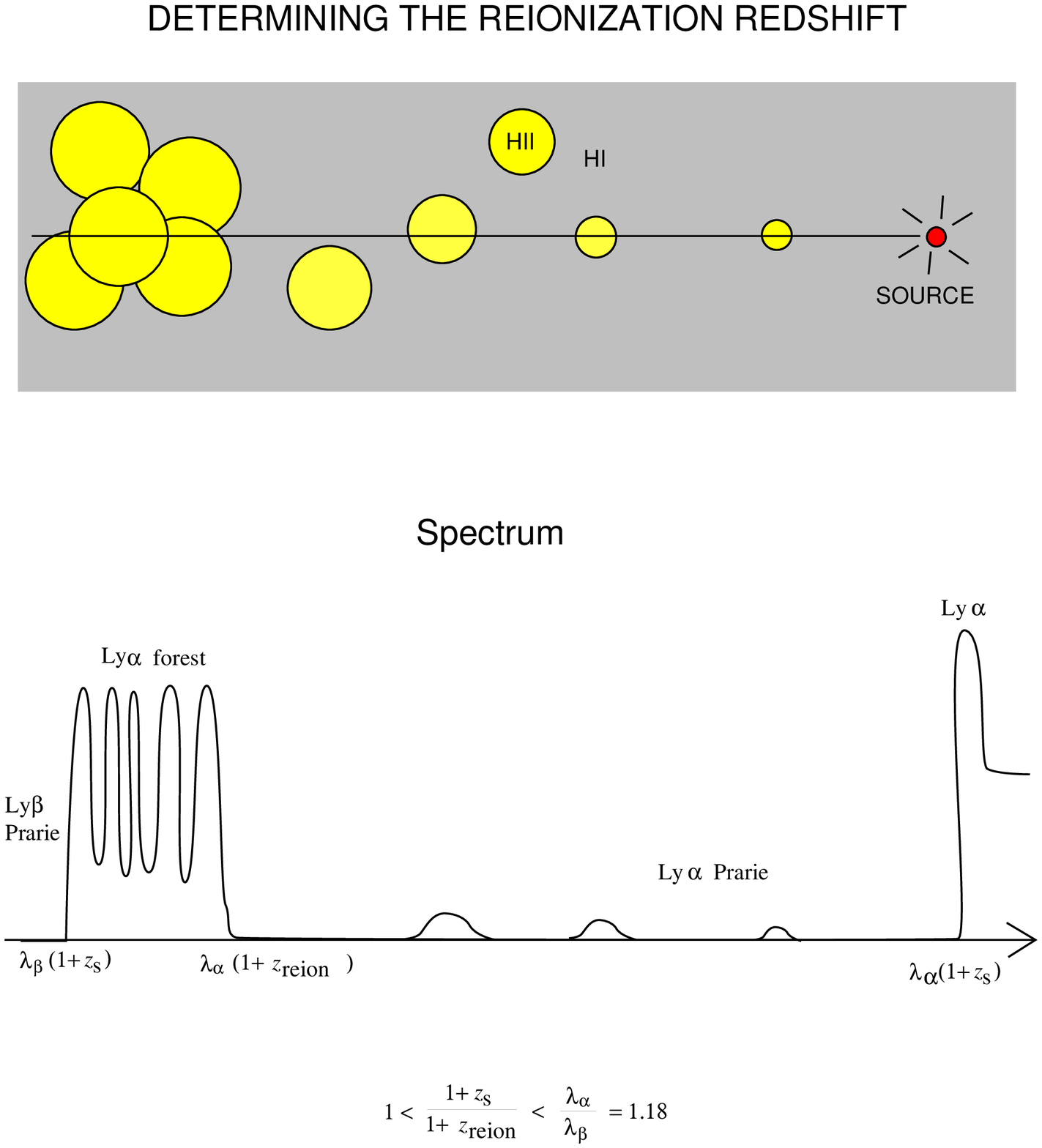}
\vspace{4.2in}
\caption{ Sketch of the expected spectrum of a source at a redshift $z_{s}$
slightly above the reionization redshift $z_{\rm reion}$. The transmitted
flux due to HII bubbles in the pre-reionization era, and the Ly$\alpha$
forest in the post-reionization era, are exaggerated for illustration.  }
\label{fig3}
\end{figure}

\subsection{Probing the Reionization Epoch and Metallicity History of the IGM}

The UV spectra of GRB afterglows can be used to measure the evolution of
the neutral intergalactic hydrogen with redshift. Figure~\ref{fig3}
illustrates schematically the expected absorption just beyond the
reionization redshift. Resonant scattering suppresses the spectrum at all
wavelengths corresponding to the Ly$\alpha$ resonance prior to
reionization.  Since the Ly$\alpha$ cross-section is very large, any
transmitted flux prior to reionization reflects a large volume of ionized
hydrogen along the line-of-sight. If the GRB is located at a redshift
larger by $>18\%$ than the reionization redshift, then the Ly$\alpha$ and
the Ly$\beta$ troughs will overlap. Unlike quasars, GRBs do not ionize the
IGM around them; their limited energy supply $\sim 10^{52}~{\rm ergs}$
\cite{WKF,Freedman,Frail} can ionize only $\sim 4\times 10^5M_\odot$ of
neutral hydrogen within their host galaxy.

Quasar spectra indicate the existence of an IGM metallicity which is a
fraction of a percent of the solar value \cite{Ellison}.  The metals were
likely dispersed into the IGM through outflows from galaxies, driven by
either supernova or quasar winds \cite{Barkana,Furlanetto}. Detection of
metal absorption lines in the spectrum of GRB afterglows, produced either
in the IGM or the host galaxy of the GRB, can help unravel the evolution of
the IGM metallicity with redshift and its link to the evolution of
galaxies. Detection of X-ray absorption by intergalactic metals can be used
to establish the existence of the warm component of the IGM which has not
been observed so far \cite{Fiore,Hellsten,Perna}.

\subsection{Cosmological Microlensing of Gamma-Ray Bursts}

Loeb \& Perna \cite{Loeb_Perna} noted the coincidence between the angular
size of a solar-mass lens at a cosmological distance and the
micro-arcsecond size of the image of a GRB afterglow. They therefore
suggested that microlensing by stars can be used to resolve the
photospheres of GRB fireballs at cosmological distances.  (Alternative
methods, such as radio scintillations, only provide a constraint on the
radio afterglow image size \cite{Goodman,WKF} but do not reveal its
detailed surface brightness distribution, because of uncertainties in the
scattering properties of the Galactic interstellar medium.)

\begin{figure}[htbp]
\includegraphics{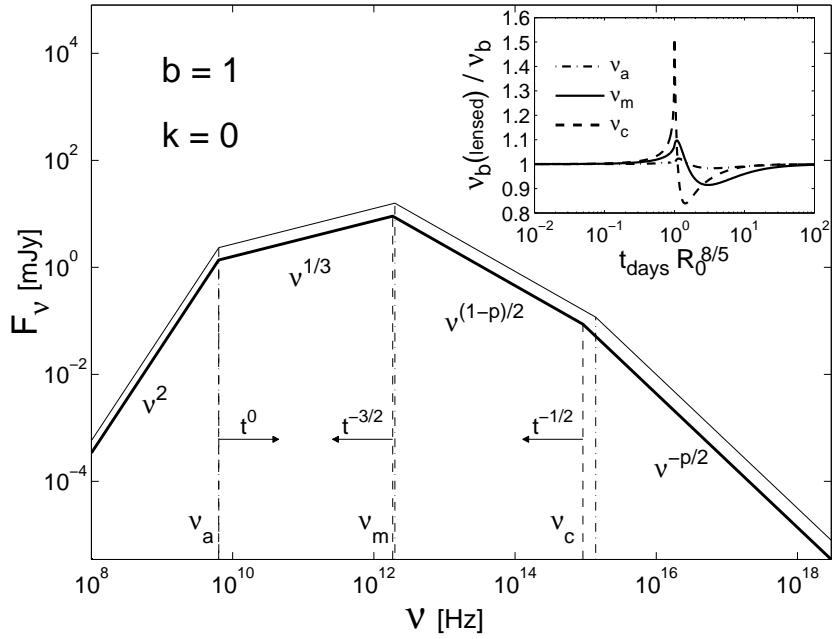}
\vspace{3.6in}
\caption{A typical broken power-law spectrum of a GRB afterglow at a
redshift $z=1$ (from Granot \& Loeb 2001). The observed flux density,
$F_\nu$, as a function of frequency, $\nu$, is shown by the boldface solid
line at an observed time $t_{\rm days}=1$ for an explosion with a total
energy output of $10^{52}~{\rm ergs}$ in a uniform interstellar medium
($k=0$) with a hydrogen density of $1~{\rm cm^{-3}}$, and post-shock energy
fractions in accelerated electrons and magnetic field of $\epsilon_e=0.1$
and $\epsilon_B=0.03$, respectively.  The thin solid line shows the same
spectrum when it is microlensed by an intervening star with an impact
parameter equal to the Einstein angle and $R_0\equiv [\theta_s(1~{\rm
day})/\theta_{\rm E}]=1$. The insert shows the excess evolution of the
break frequencies $\nu_{\rm b}=\nu_a,~\nu_m$ and $\nu_c$ (normalized by
their unlensed values) due to microlensing. }
\label{fig4}
\end{figure}

The fireball of a GRB afterglow is predicted to appear on the sky as a ring
(in the optical band) or a disk (at low radio frequencies) which expands
laterally at a superluminal speed, $\sim \Gamma c$, where $\Gamma\gg1$ is
the Lorentz factor of the relativistic blast wave which emits the afterglow
radiation \cite{Waxman,Sari,Panaitescu,Granot}.  For a spherical explosion
into a constant density medium (such as the interstellar medium), the
physical radius of the afterglow image is of order the fireball radius over
$\Gamma$, or more precisely \cite{Granot}
\begin{equation}
R_{\rm s}= 3.9\times 10^{16}\left({E_{52}\over n_1}\right)^{1/8}
\left({t_{\rm days}\over 1+z}\right)^{5/8}~{\rm cm},
\label{eq:r_s}
\end{equation}
where $E_{52}$ is the hydrodynamic energy output of the GRB explosion in
units of $10^{52}~{\rm ergs}$, $n_1$ is the ambient medium density in units
of $1~{\rm cm^{-3}}$, and $t_{\rm days}$ is the observed time in days.  At
a cosmological redshift $z$, this radius of the GRB image occupies an angle
$\theta_{\rm s}=R_{\rm s}/D_A$, where $D_A(z)$ is the angular diameter
distance at the GRB redshift, $z$. For the typical cosmological distance,
$D_A\sim 10^{28}~{\rm cm}$, the angular size is of order a micro-arcsecond
($\mu$as).  Coincidentally, this image size is comparable to the Einstein
angle of a solar mass lens at a cosmological distance,
\begin{equation}
\theta_{\rm E}=\left({4GM_{\rm lens}\over c^2 D}\right)^{1/2}= 1.6
\left({M_{\rm lens}\over 1 M_\odot}\right)^{1/2}\left({D\over 10^{28}~{\rm
cm}}\right)^{-1/2}~\mu{\rm as},
\label{eq:1}
\end{equation}
where $M_{\rm lens}$ is the lens mass, and $D\equiv {D_{\rm os}D_{\rm ol}/
D_{\rm ls}}$ is the ratio of the angular-diameter distances between the
observer and the source, the observer and the lens, and the lens and the
source \cite{Schneider}.  Loeb \& Perna \cite{Loeb_Perna} showed that
because the ring expands laterally faster than the speed of light, the
duration of the microlensing event is only a few days rather than tens of
years, as is the case for more typical astrophysical sources which move at
a few hundred ${\rm km~s^{-1}}$ or $\sim 10^{-3}c$.

The microlensing lightcurve goes through three phases: (i) constant
magnification at early times, when the source is much smaller than the
source-lens angular separation; (ii) peak magnification when the ring-like
image of the GRB first intersects the lens center on the sky; and (iii)
fading magnification as the source expands to larger radii.

\begin{figure}[htbp]
\includegraphics{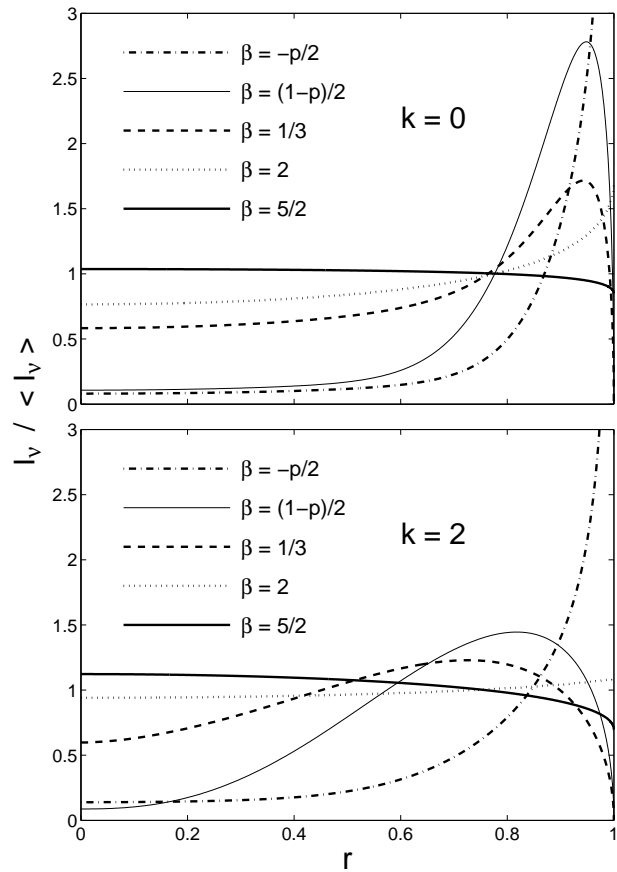}
\vspace{4.8in}
\caption{The surface brightness, normalized by its average value, as a
function of the normalized radius, $r$, from the center of a GRB afterglow
image (where $r=0$ at the center and $r=1$ at the outer edge). The image
profile changes considerably between different power-law segments of the
afterglow spectrum, $F_{\nu}\propto\nu^{\beta}$ (see
Figure~\ref{fig4}). There is also a strong dependence on the power--law
index of the radial density profile of the external medium around the
source, $\rho\propto R^{-k}$ (taken from Granot \& Loeb 2001).  }
\label{fig5}
\end{figure}

Granot \& Loeb \cite{Granot_Loeb} calculated the radial surface brightness
profile (SBP) of the image of a Gamma-Ray-Burst (GRB) afterglow as a
function of frequency and ambient medium properties, and inferred the
corresponding magnification lightcurves due to microlensing by an
intervening star. The afterglow spectrum consists of several power-law
segments separated by breaks, as illustrated by Figure~\ref{fig4}. The
image profile changes considerably across each of the spectral breaks, as
shown in Figure~\ref{fig5}. It also depends on the power--law index, $k$,
of the radial density profile of the ambient medium into which the GRB
fireball propagates.  Gaudi \& Loeb \cite{Gaudi_Loeb} have shown that
intensive monitoring of a microlensed afterglow lightcurve can be used to
reconstruct the parameters of the fireball and its environment.  The
dependence of the afterglow image on frequency offers a fingerprint that
can be used to identify a microlensing event and distinguish it from
alternative interpretations.  It can also be used to constrain the
relativistic dynamics of the fireball and the properties of its gaseous
environment. At the highest frequencies, the divergence of the surface
brightness near the edge of the afterglow image ($r=1$ in
Figure~\ref{fig5}) depends on the thickness of the emitting layer behind
the relativistic shock front, which is affected by the length scale
required for particle acceleration and magnetic field amplification behind
the shock\cite{Medvedev,Gruzinov}.

Ioka \& Nakamura \cite{Ioka} considered the more complicated case where the
explosion is collimated and centered around the viewing axis. More general
orientations that violate circular symmetry need to be considered in the
future.

\begin{figure}[htbp]
\includegraphics{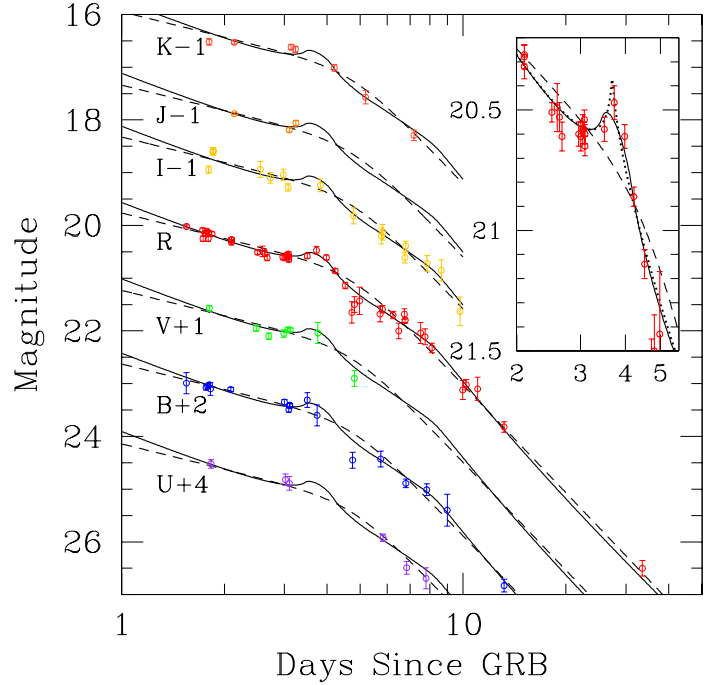}
\vspace{3.9in}
\caption{ $UBVRIJK$ photometry of GRB 000301C as a function of time in days
from the GRB trigger (from Garnavich et al. 2000; Gaudi et al.  2001).
Data points have been offset by the indicated amount for clarity.  The
dashed line is the best-fit smooth, double power-law lightcurve (with no
lensing), while the solid line is the overall best-fit microlensing model,
where the SBP has been determined from direct inversion.  The inset shows
the $R$-band data only.  The dotted line is the best-fit microlensing model
with theoretically calculated SBP, for $k=0$ and $\nu>\nu_c$.  }
\label{fig6}
\end{figure}

\subsubsection{GRB 000301C}

Garnavich, Loeb, \& Stanek \cite{Garnavich} have reported the possible
detection of a microlensing magnification feature in the optical-infrared
light curve of GRB 000301C (see Figure~\ref{fig6}). The achromatic
transient feature is well fitted by a microlensing event of a $0.5 M_\odot$
lens separated by an Einstein angle from the source center, and resembles
the prediction of Loeb \& Perna \cite{Loeb_Perna} for a ring-like source
image with a narrow fractional width ($\sim 10\%$). Alternative
interpretations relate the transient achromatic brightening to a higher
density clump into which the fireball propagates \cite{Berger}, or to a
refreshment of the decelerating shock either by a shell which catches up
with it from behind or by continuous energy injection from the source
\cite{Zhang}. However, the microlensing model has a smaller number of free
parameters. If with better data, a future event will show the generic
temporal and spectral characteristics of a microlensing event, then these
alternative interpretations will be much less viable.  A galaxy
2$^{\prime\prime}$ from GRB~000301C might be the host of the stellar lens,
but current data provides only an upper-limit on its surface brightness at
the GRB position. The existence of an intervening galaxy increases the
probability for microlensing over that of a random line-of-sight.

\begin{figure}[htbp]
\includegraphics{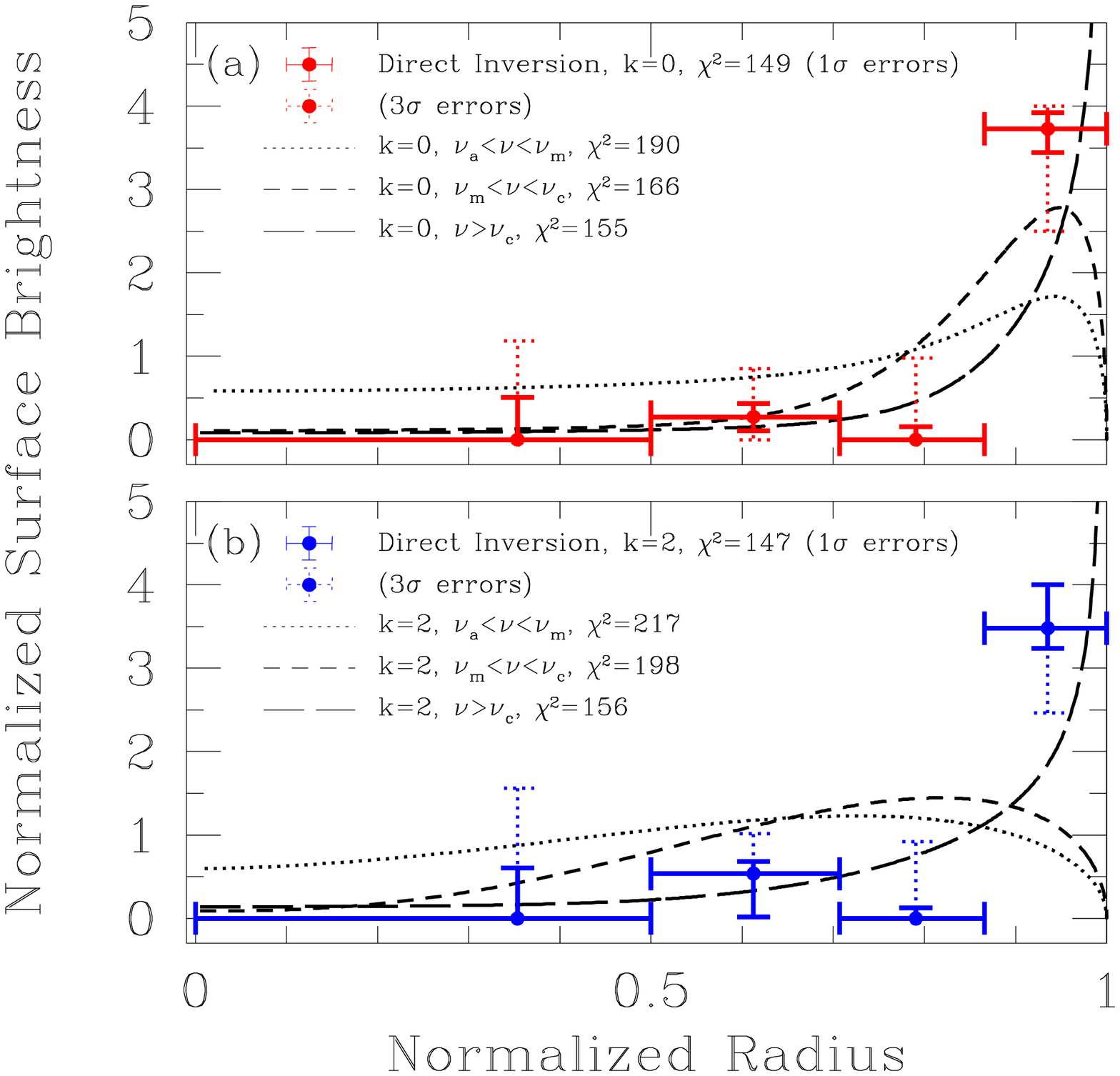}
\vspace{3.9in}
\caption{Fitting GRB 000301C with different SBPs as a function of
normalized radius (taken from Gaudi et al. 2001).  The points are the SBPs
determined from direct inversion, with $1\sigma$ errors (solid) and
$3\sigma$ errors (dotted).  The curves are theoretically calculated SBPs
for various frequency regimes (see Figure~\ref{fig5}).  (a) Uniform
external medium, $k=0$.  (b) Stellar wind environment, $k=2$. The number of
degrees of freedom is 92 for the direct inversion points and 89 for the
curves.}
\label{fig7}
\end{figure}

Gaudi, Granot, \& Loeb \cite{GGL} have shown that direct inversion of the
observed light curve for GRB 000301C yields a surface brightness profile
(SBP) of the afterglow image which is strongly limb-brightened, as expected
theoretically (see Figure~\ref{fig7}).

Obviously, realistic lens systems could be more complicated due to the
external shear of the host galaxy or a binary companion.  Mao \& Loeb
\cite{Mao} calculated the magnification light curves in these cases, and
found that binary lenses may produce multiple peaks of magnification.  They
also demonstrated that {\it all} afterglows are likely to show variability
at the level of a few percent about a year following the explosion, due to
stars which are separated by tens of Einstein angles from the
line-of-sight.

{\it What is the probability for microlensing?} If the lenses are not
strongly clustered so that their cross-sections overlap on the sky, then
the probability for having an intervening lens star at a projected angular
separation $\theta$ from a source at a redshift $z\sim 2$ is~~$\sim 0.3
\Omega_\star (\theta/\theta_E)^2$ \cite{Press,Blaes,Nemiroff,Nemiroff2},
where $\Omega_\star$ is the cosmological density parameter of stars.  The
value of $\Omega_\star$ is bounded between the density of the luminous
stars in galaxies and the total baryonic density as inferred from Big Bang
nucleosynthesis, $7\times 10^{-3} < \Omega_\star< 5\times 10^{-2}$
\cite{Fukugita}. Hence, {\it all} GRB afterglows should show evidence for
events with $\theta \sim 30\theta_E$, for which microlensing provides a
small perturbation to the light curve\cite{Mao}. (This crude estimate
ignores the need to subtract those stars which are embedded in the dense
central regions of galaxies, where macrolensing dominates and the
microlensing optical depth is of order unity.) However, only one out of
roughly a hundred afterglows is expected to be strongly microlensed with an
impact parameter smaller than the Einstein angle. Indeed, Koopmans \&
Wambsganss \cite{Koopmans} have found that the `a posteriori' probability
for the observed microlensing event of GRB 000301C along a random
line-of-sight is between 0.7--2.7\% if 20-100\% of the dark matter is in
compact objects.

Microlensing events are rare but precious.  Detailed monitoring of a few
strong microlensing events among the hundreds of afterglows detected per
year by the forthcoming Swift satellite, could be used to constrain the
environment and the dynamics of relativistic GRB fireballs, as well as
their magnetic structure and particle acceleration process.

\section{Illumination of the Intergalactic Medium by Quasars}

\subsection{Ly$\alpha$ Halos}

The absorption trough created in the spectrum of a distant sources due to
neutral hydrogen diminishes down to undetectable flux levels as soon as the
mean neutral fraction of hydrogen is larger than $\sim 10^{-4}$ (see
equation~\ref{eq:G-P}). Once the transmitted flux reaches very low levels,
it is difficult to infer whether the neutral fraction of hydrogen is as low
as 0.01\% or as high as 100\%.  A novel way to study the reionization phase
transition relies on a direct detection of intergalactic neutral
hydrogen. Loeb \& Rybicki \cite{Loeb_Rybicki,Rybicki} have shown that the
neutral gas prior to reionization can be probed through narrow-band imaging
of embedded Ly$\alpha$ sources. The physical situation is analogous to the
appearance of a halo of scattered light around a street lamp which is
embedded in a dense fog. The ``street lamp'' in this metaphor is a
high--redshift quasar which emits Ly$\alpha$ photons into the surrounding
neutral hydrogen gas.  The IGM scatters these photons and acts as a fog.

The radiation of the first galaxies is strongly absorbed shortward of their
restframe Ly$\alpha$ wavelength by neutral hydrogen in the intervening IGM.
However, the Ly$\alpha$ photons emitted by these sources are not eliminated
but rather scatter until they redshift out of resonance and escape due to
the Hubble expansion of the surrounding intergalactic hydrogen (see
Fig. 8).  Detection of the diffuse Ly$\alpha$ halos around high redshift
sources would provide a unique tool for probing the neutral IGM before the
epoch of reionization. Loeb \& Rybicki \cite{Loeb_Rybicki,Rybicki} explored
the above effect for a uniform, fully-neutral IGM in a pure Hubble flow.
It is important to extend this analysis to more realistic cases of sources
embedded in an inhomogeneous IGM, which is partially ionized by these
sources \cite{Bromm2}.  It would be interesting to extract particular
realizations of the perturbed IGM around massive galaxies from numerical
simulations, and to apply a suitable radiative transfer code in propagating
the Ly$\alpha$ photons from the embedded galaxies. Mapping of the
properties of the associated Ly$\alpha$ halos will allow to assess their
detectability with future observations.

\begin{figure}[htbp]
\includegraphics{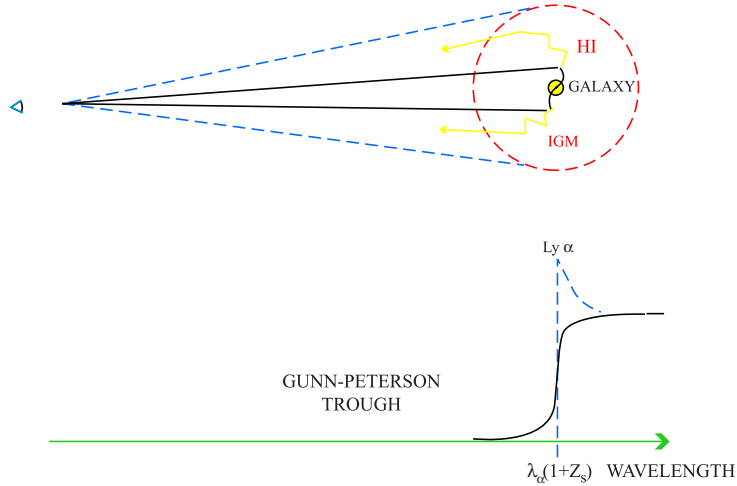}
\vspace{3.05in}
\caption{{\it Loeb--Rybicki halos:}\/ Scattering of \lya line photons
from a galaxy embedded in the neutral intergalactic medium prior to
reionization. The line photons diffuse in frequency due to the Hubble
expansion of the surrounding medium and eventually redshift out of
resonance and escape to infinity.  A distant observer sees a \lya halo
surrounding the source, along with a characteristically asymmetric
line profile. The observed line should be broadened and redshifted by
about one thousand ${\rm km~s^{-1}}$ relative to other lines (such as
H$_\alpha$) emitted by the galaxy.  }
\label{fig8}
\end{figure}

\subsection{Magnetization of the Intergalactic Medium by Quasar Outflows}

Magnetic fields and cosmic rays play an important dynamical role in the
interstellar medium of galaxies. However, they are often ignored in
discussions of the intergalactic medium (IGM). {\it How significant is the
pressure from these non-thermal components in the IGM? Has the magnetic
field observed in collapsed objects, such as galaxies or clusters of
galaxies, originated from the IGM?}  The first question is particularly
relevant for the reconstruction of the mass distribution of galaxy clusters
from X-ray data.

\begin{figure}[htbp]
\includegraphics{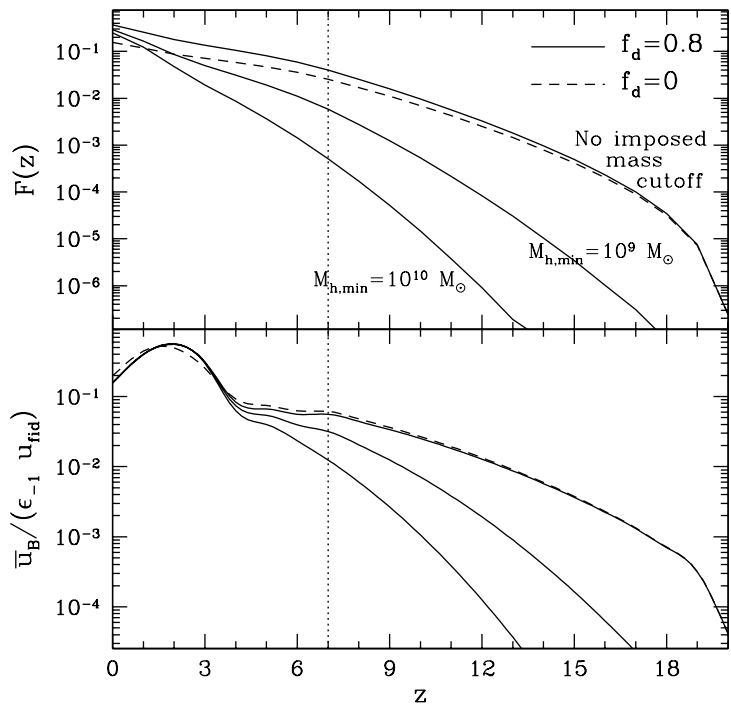}
\vspace{3.9in}
\caption{{\it Upper Panel:} Volume filling fraction of magnetized bubbles
$F(z)$, as a function of redshift (taken from Furlanetto \& Loeb 2001).
{\it Lower Panel:} Ratio of magnetic energy density,
$\bar{u}_B/\epsilon_{-1}$, to a fiducial thermal energy density $u_{fid} =
3 n(z) k T_{IGM}$ of a photoionized IGM, where $T_{IGM} = 10^4 \kel$, as a
function of redshift.  The solid curves assume a different minimum mass for
a galaxy out of which a quasar outflow may originate, $M_{h,min}$. Examples
include a case where this minimum mass is determined by atomic cooling
before reionization and by infall suppression afterward (top curve), and
fixed-value cases with $M_{h,min} = 10^9 \msun$ (middle curve), and
$M_{h,min} = 10^{10} \msun$ (bottom curve). The vertical dotted line
indicates the assumed redshift of reionization, $z_r=7$.}
\label{fig9}
\end{figure}

Outflows from quasars inevitably pollute the IGM with magnetic fields. The
short-lived activity of a quasar leaves behind an expanding magnetized
bubble in the IGM. Furlanetto \& Loeb \cite{Furlanetto} modeled the
expansion of the remnant quasar bubbles and calculated their distribution
as a function of size and magnetic field strength at different redshifts.
They have found that generically by a redshift $z\sim 3$, about 5--$20\%$
of the IGM volume is filled by magnetic fields with an energy density $\ga
10\%$ of the mean thermal energy density of a photo-ionized IGM at $\sim
10^4$ K (see Figure~\ref{fig9}). As massive galaxies and X-ray clusters
condense out of the magnetized IGM, the adiabatic compression of the
magnetic field could result in the field strength observed in these systems
without a need for further dynamo amplification.  The intergalactic
magnetic field could also provide a non--thermal contribution to the
pressure of the photo-ionized gas that may account for the claimed
discrepancy between the simulated and observed Doppler width distributions
of the Ly$\alpha$ forest at $z\ga 2$ \cite{Bryan,Theuns}. However, the
supplied magnetic energy is unlikely to be dynamically important today
since the present-day IGM was heated by gravitationally-induced shocks to
an average mass-weighted temperature of $\sim 3\times 10^6$K \cite{Dave},
larger by two orders of magnitude than the magnetic energy input from
quasars.

\subsection{Particle Acceleration in Magnetized Intergalactic Shocks}

Even though the present-day magnetic pressure may not be dynamically
important, the magnetization of the IGM by quasar outflows has important
consequences.  Once magnetized, the shocks produced in the IGM during the
formation of large-scale structure can accelerate a population of highly
relativistic electrons by the Fermi mechanism, similarly to the
collisionless shocks of supernova remnants. The accelerated electrons emit
synchrotron radiation as they gyrate in the embedded magnetic fields and
produce $\gamma$-ray radiation by inverse--Compton scattering the microwave
background photons. This non-thermal radiation spans some $\sim 20$ orders
of magnitude in photon energies from the radio to the TeV $\gamma$-ray
regime. The radiation is expected to delineate the strong shocks formed in
the cosmic web of large scale sheets and filaments in the IGM.  The
brightest emission originates from the strongest intergalactic shocks
around galaxy clusters.

\subsection{Non-Thermal Radio and Gamma-Ray Emission}

More than a third of all X-ray clusters with luminosities $\ga 10^{45}~{\rm
erg~s^{-1}}$ possess diffuse radio halos \cite{Giovannini}.  Based on
energy arguments and circumstantial evidence, these radio halos are
believed to be caused by synchrotron emission from shock-accelerated
electrons during the merger events of their host clusters
(\cite{Harris,Tribble,Feretti}; see \cite{Liang} for references to
alternative, less successful models).  These highly-relativistic electrons
cool primarily through inverse-Compton scattering off the microwave
background. Since their cooling time is much shorter than the dynamical
time of their host cluster, the radio emission is expected to last as long
as the shock persists and continues to accelerate fresh electrons to
relativistic energies. Intergalactic shocks also occur along the filaments
and sheets that channel mass into the clusters. These structures, also
traced by the galaxy distribution \cite{Doroshkevich}, are induced by
gravity and form due to converging large-scale flows in the IGM.

\begin{figure}[t]
\includegraphics{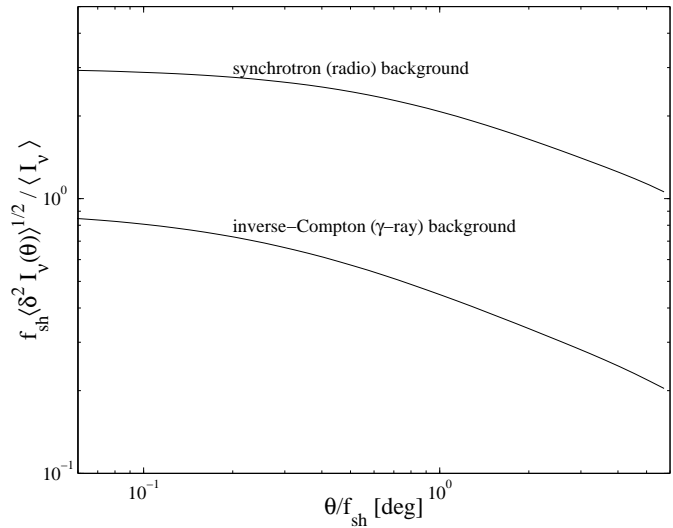}
\vspace{2.9in}
\caption{Fractional intensity fluctuations, $(\langle
I_\nu(0)I_\nu(\theta)\rangle-\langle I_\nu\rangle^2)^{1/2}/ \langle I_\nu
\rangle$, in synchrotron (radio) and inverse-Compton ($\gamma$-ray)
background flux (taken from Waxman \& Loeb 2000).  The dimensionless
coefficient, $f_{\rm sh}$, is of order unity. The ratio of synchrotron to
CMB intensity is $\langle I^{\rm syn}_\nu\rangle/I^{\rm
CMB}_\nu=6\times10^{-6} f_{\rm sh}^{-2}(\xi_B/0.01)(\nu/{10\rm GHz})^{-3}$,
where the post-shock magnetic energy fraction $\xi_B$ is assumed to be of
order 0.01, corresponding to a magnetic field strength of 0.1$\mu$G in the
virialization shock of X-ray clusters.  The post-shock energy fraction
carried by relativistic electrons is assumed to be $\xi_e\sim 0.05$,
similar to the value inferred in the shocks of supernova remnants.}
\label{fig10}
\end{figure}

The intergalactic synchrotron emission contaminates the cosmic microwave
background (CMB) anisotropies -- relic from the epoch of recombination, and
as such needs to be considered in the design and analysis of anisotropy
experiments at low frequencies.  Previous estimates of synchrotron
contamination of the cosmic anisotropies focused on Galactic emission,
which occurs primarily at low Galactic latitudes and large angular scales
\cite{Tegmark2}.  Waxman \& Loeb \cite{Waxman_Loeb} calculated the
intergalactic synchrotron contribution to the fluctuations in the radio sky
as a function of frequency and angular scale (see Figure~\ref{fig10}). They
assumed that most of the emission originates from the virialization shocks
around X-ray clusters, and used the Press--Schechter \cite{PS} mass
function to describe the abundance of such clusters as a function of
redshift (see also subsequent work in \cite{Fujita}).  Although the
synchrotron background amounts to only a small fraction of the CMB
intensity, Waxman \& Loeb \cite{Waxman_Loeb} showed that its fluctuations
could dominate over the primordial CMB fluctuations at low photon
frequencies, $\nu \la 10$~GHz.  They found that radio emission from cluster
shocks contributes a fluctuation amplitude of $\sim 40 \mu$K $\times
(\nu/10{\rm GHz})^{-3}$ to the CMB on angular scales between 1 and
$0.1^\circ$, respectively.  Interestingly, current anisotropy experiments
are just sensitive to this level of fluctuations.
%\footnote{See table summary of current experiments at
%http://www.hep.upenn.edu/$\~$max/index.html, or at
%http://cfa-www.harvard.edu/$\~$mwhite/cmbexptlist.html}.  
Existing detections by CAT ($50\pm15\mu$K at 15~GHz on 0.2--0.5$\arcdeg$
scales) and OVRO ($56^{+8.5}_{-6.6}\mu$K at 20~GHz on 0.1--0.6$\arcdeg$
scales), as well as 95\% upper limits ($\la 40\mu$K on arcminute scales at
9-15 GHz by the ATCA, RYLE and VLA detectors) are consistent with our
prediction.
%\footnote{Note that although the cluster contribution declines
%rapidly at high frequencies, some of these experiments, like CAT or OVRO,
%measured fluctuations at only one frequency and could not reject a
%synchrotron contribution to the measured signal.}.

Loeb \& Waxman \cite{Loeb_Waxman} have shown that a significant fraction of
the diffuse $\gamma$--ray background \cite{Sreekumar} might have been
generated by the shocks resulting from the formation of large-scale
structure in the intergalactic medium. Similarly to the collisionless
shocks of supernova remnants \cite{Blandford}, these shocks produce a
population of highly-relativistic electrons with a maximum Lorentz factor
$\ga 10^7$ that scatter a small fraction of the microwave background
photons in the present-day universe up to $\gamma$-ray energies, thereby
providing the $\gamma$-ray background.  The predicted flux agrees with the
observed diffuse background over more than four decades in photon energy
and is not sensitive to the precise magnetic field value, provided that the
fraction of shock energy carried by relativistic electrons is
$\xi_e\sim0.05$ (a value consistent with that inferred for supernovae
shocks).  The same electrons that emit $\gamma$-rays by inverse-Compton
scattering of microwave background photons, also produce synchrotron
radiation in the radio band due to intergalactic magnetic fields. The
existence of magnetic fields with an amplitude $\ga 0.1 \mu$G, is inferred
in cluster halos \cite{Kim,FuscoFemiano,Rephaeli,Kaastra}, and is also
required for the Fermi acceleration of these electrons.  The appearance of
radio halos around young X-ray clusters is therefore a natural consequence
and an important test of this model for the extragalactic $\gamma$-ray
background. The combination of radio and $\gamma$-ray data can be used to
calibrate $\xi_e$ and determine the strength of the intergalactic magnetic
field.

The assumed acceleration of electrons by collisionless shocks is similar to
that observed in supernova remnants.  Recent X-ray \cite{Koyama1,Koyama2}
and TeV \cite{Tanimori,Muraishi} observations of the supernova remnants
SN1006 and SNR RX J1713.7--3946 imply that electrons are accelerated in the
remnant shocks up to an energy $\sim 100$~TeV, and are confined to the
collisionless fluid by magnetic fields.  These shocks have a velocity of
order $10^3~{\rm km~s^{-1}}$, similar to the velocity of the intergalactic
shocks we consider here.  Although the plasma density is very different in
the two cases, the density may be scaled out of the problem by measuring
time in units of $\omega_{pe}^{-1}$, where $\omega_{pe}$ is the electron
plasma frequency (shock characteristics may also depend on the pre-shock
magnetic field, which introduces a dimensionless parameter into the
problem, $\omega_{ce}/\omega_{pe}$, where $\omega_{ce}$ is the electron
cyclotron frequency. However, for both supernova and IGM shocks,
$\omega_{ce}/\omega_{pe}\ll1$, suggesting a similar behavior in both
cases).  For supernova shocks, the inferred energy density in relativistic
electrons constitutes $1$--$10\%$ of the post-shock energy density in these
remnants \cite{Ellison2}, a fraction consistent with the global ratio
between the mean energy density of cosmic-ray electrons and the turbulent
energy density in the interstellar medium of our galaxy.

The production of anisotropic backgrounds of radio and $\gamma$-ray
radiation by strong intergalactic shocks is a natural consequence of
structure formation in the Universe. The brightest emission originates from
the shocks on Mpc scales around newly formed, massive X-ray clusters.  The
foreground synchrotron fluctuations might be comparable to the anisotropy
signals detected by existing low-frequency microwave background
experiments, and can be easily isolated through multi-frequency
observations. Polarization anisotropy experiments could then constrain the
coherence length of the intergalactic magnetic field.

Different astrophysical sources may contribute to the strength of the
intergalactic magnetic fields; aside from radio sources
\cite{Daly90,Furlanetto} they include: the large scale shocks themselves
\cite{Kulsrud97}, the first generation of stars \cite{Rees87} and
supernova-driven winds from galaxies \cite{Kronberg99}.  Different
contributions may lead to different scalings of the magnetic field energy
density with IGM gas parameters.  The non-thermal $\gamma$-ray emission is
only weakly dependent on the magnetic field strength
\cite{Loeb_Waxman}. However, the non-thermal radio emission is highly
sensitive to this parameter and different models may lead to a different
dependence of the radio emission on scale and redshift, which may have
observable consequences (e.g. modifying the functional dependence of the
correlation function shown in Figure \ref{fig10}).

Recent observations of the giant radio galaxy NGC 315 may constitute the
first direct detection of collisionless large scale structure shocks,
accelerating electrons to high energies \cite{Ensslin01}.  Radio
observations of this source are most naturally explained by the existence
of a large scale shock produced by a flow converging towards an
intersection of galaxy filaments \cite{Ensslin01}.

Waxman \& Loeb \cite{Waxman_Loeb} predicted a fluctuation amplitude $\ga
40\%$ in the $\gamma$-ray background intensity on sub-degree scale, and the
existence of extended, $\ga 1\arcdeg$, $\gamma$-ray halos, associated with
newly formed massive clusters. On scales larger than a degree the
fluctuation amplitude declines and is well below the anisotropy limits from
EGRET (see Figure 5 in \cite{Sreekumar}).  Detection of the predicted
signals will provide a calibration of the uncertain model parameter,
$\xi_e$.  The high-energy maps required to detect the predicted anisotropy
signal will be made between 20 MeV and 300 GeV by the GLAST instrument
(planned for launch in 2005; see http://glast.gsfc.nasa.gov/), which is
expected to be more sensitive than EGRET by an order-of-magnitude
\cite{Bloom96}.  The predicted $\gamma$-ray halos may constitute a
significant fraction of the unidentified extra-Galactic EGRET sources
\cite{Totani1,Totani2}.  However, since the angular extension of the
brightest halos is large, a more careful analysis is required in order to
assess the detectability of such halos by EGRET.

A future, dedicated, all-sky anisotropy experiment, operating at several
frequencies below 10 GHz, would be able to map the fluctuations in the
intergalactic synchrotron background.  The resulting synchrotron map could
then be cross-correlated with full-sky maps at hard X-ray or $\gamma$-ray
energies to confirm its cosmic origin. Identification of the synchrotron
fluctuations together with their counterpart inverse-Compton emission of
hard X-rays or $\gamma$-rays by the same population of shock-accelerated
electrons, can be used to empirically determine the strength and spatial
distribution of the intergalactic magnetic field.  Similarly, the
correlation between radio and $\gamma$-ray halos may be detectable around
individual X-ray clusters.  Strong radio halos could be the best indicators
for bright $\gamma$-ray clusters, which would provide the first obvious
targets for GLAST.

Semi-analytic models \cite{Waxman_Loeb} identify the intergalactic shocks
with smooth spherical accretion of gas onto clusters, while in reality they
result from asymmetric mergers as well as from converging flows in large
scale sheets and filaments. The more realistic emission from these complex
geometries can be best modeled through detailed hydrodynamic simulations
\cite{Miniati1,Miniati3,Keshet}. Mergers of comparable mass clusters, for
example, would tend to produce only mild shocks due to the prior heating of
the shocked gas, leading to reduced non-thermal emission due to the steep
power-law slope of the accelerated electrons.

In analogy with the collisionless shocks of supernova remnants, the
intergalactic shocks are also expected to accelerate a hadronic cosmic-ray
component that would acquire a fraction ($\sim 10$--$50\%$) of the
post-shock energy more substantial than that carried by relativistic
electrons ($\sim 1$--$10\%$, see \cite{Ellison2}). Collisions of the
hadronic cosmic-rays with IGM protons produce pions, $\pi^0$, which decay
into $\gamma$-ray photons; however for typical parameters this radiation
component is expected to be sub-dominant relative to the inverse-Compton
production of $\gamma$-rays by the relativistic electrons
\cite{Loeb_Waxman,Waxman_Loeb,Miniati2}.  The buoyancy of the relativistic
cosmic--ray fluid may lead to instabilities that would eventually separate
it from the IGM gas and allow it to expand into the large-scale voids that
fill most of the volume in between the sheets and filaments of the IGM. For
a relativistic fluid, the adiabatic decrease of the pressure with
increasing volume ($p\propto V^{-4/3}$) is milder than for the IGM gas
($p\propto V^{-5/3}$). It is possible that the intergalactic voids are
currently being dominated by cosmic--ray and magnetic field pressure which
suppresses the formation of galaxies there.

\bigskip
\noindent{\bf Acknowledgements}

\noindent
I thank all my collaborators on the topics described in this brief review:
Rennan Barkana, Benedetta Ciardi, Steve Furlanetto, Peter Garnavitch, Scott
Gaudi, Jonathan Granot, Zoltan Haiman, Shude Mao, Rosalba Perna, George
Rybicki, Kris Stanek, and Eli Waxman. This work was supported in part by
NASA grants NAG 5-7039, 5-7768, and by NSF grants AST-9900877, AST-0071019.

\vfil\eject

\end{document}